\documentclass{article}
\usepackage[utf8]{inputenc}

\title{Soft Metamaterials: Adaptation and Intelligence}

\author{Colin Scheibner $^{1,2}$, Michel Fruchart $^{1,2}$, Vincenzo Vitelli $^{1,2,3}$} 
\date{\footnotesize 
$^1$ James Franck Institute, The University of Chicago, Chicago, IL 60637, USA\\
$^2$ Department of Physics, The University of Chicago, Chicago, IL 60637, USA\\
$^3$ Kadanoff Center for Theoretical Physics, The University of Chicago, Chicago, IL 60637, USA
}

\usepackage{graphicx}
\usepackage{xcolor}
\input{tokcycle.sty}

\newcount\wordcount
\newcount\lettercount
\newcount\wordlimit
\newif\ifinword
\newif\ifrunningcount
\newif\ifsummarycount
\def\limitcolor{red}
\catcode`@=11
\def\addtomacro#1#2{\tc@defx#1{#1#2}}
\def\changecolor#1{\tctestifx{.#1}{}{\addcytoks{\color{#1}{}}%
  \tc@defx\countemcurrentcolor{#1}}}
\catcode`@=12
\def\dumpword{%
  \addcytoks[1]{\accumword}%
  \ifinword\global\advance\wordcount 1\relax
    \ifrunningcount\addcytoks[x]{$^{\the\wordcount,\the\lettercount}$}\fi
    \ifnum\wordcount=\wordlimit\relax\changecolor{\limitcolor}\fi
  \fi%
  \inwordfalse
  \def\accumword{}}
\def\addletter#1{%
  \tctestifcatnx A#1{\global\advance\lettercount 1\relax\inwordtrue}{\dumpword}%
  \addtomacro\accumword{#1}}
\xtokcycleenvironment\countem
  {\addletter{##1}}
  {\dumpword\groupedcytoks{\processtoks{##1}\dumpword\expandafter}\expandafter
    \changecolor\expandafter{\countemcurrentcolor}}
  {\dumpword\addcytoks{##1}}
  {\dumpword\addcytoks{##1}}
  {\stripgroupingtrue\def\accumword{}\def\countemcurrentcolor{.}
    \global\wordcount=0\relax\global\lettercount 0\relax}
  {\dumpword\ifsummarycount\tcafterenv{%
    \par(Wordcount=\the\wordcount, Lettercount=\the\lettercount)}\fi}

\definecolor{tab_blue}{HTML}{1F77B4}
\definecolor{tab_orange}{HTML}{FF7F0E}
\definecolor{tab_green}{HTML}{2CA02C}
\definecolor{tab_red}{HTML}{D62728}
\definecolor{tab_purple}{HTML}{9467BD}
\definecolor{tab_brown}{HTML}{8C564B}
\definecolor{tab_pink}{HTML}{E377C2}
\definecolor{tab_gray}{HTML}{7F7F7F}
\definecolor{tab_olive}{HTML}{BCBD22}
\definecolor{tab_cyan}{HTML}{17BECF}

\begin{document}

\maketitle

\noindent \textbf{Status.}
\countem 

\begin{figure}
    \centering
    \includegraphics[width=0.7 \textwidth ]{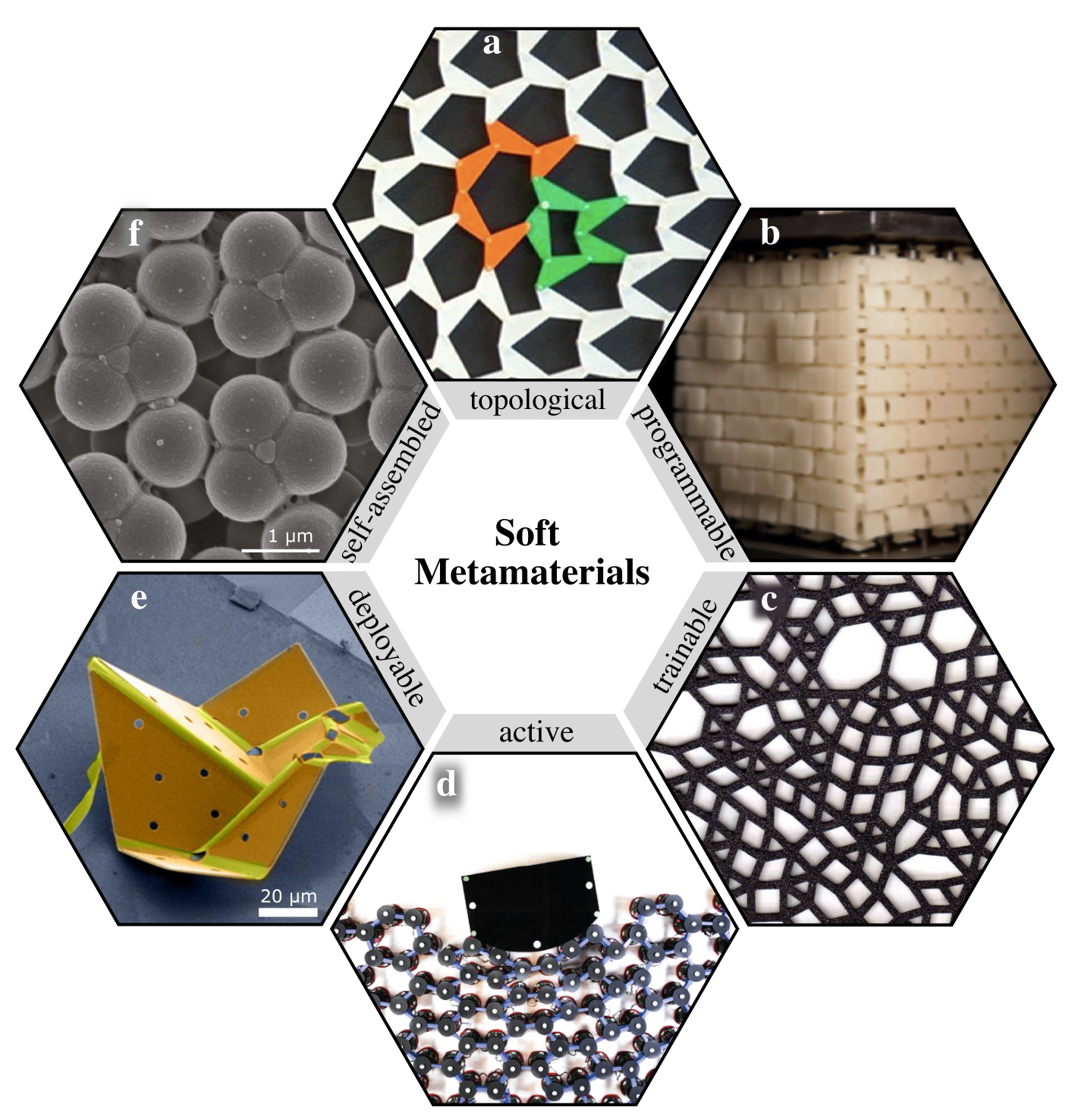}
    \caption{
    \label{Fig1}
    \textbf{Soft metamaterials.} {\bf a.}~A topological metamaterial exhibiting a robust soft mode bound to a crystalline defect, adapted from Paulose,~et~al.~Nat.~Phys.~(2015). 
    {\bf b.}~A programmable metamaterial using multistability to create a desired post-buckling form, adapted from Coulais,~et~al.~Nature~(2016). 
    {\bf c.}~A disordered network of foam bonds can be trained via repeated compression and directed aging, adapted from Pashine,~et~al.~Sci.~Adv.~(2019). 
    {\bf d.}~An active metamaterial consisting of robotic building blocks with distributed energy sources, adapted from Brandenbourger,~et~al.~arXiv:2108.08837~(2022). 
    {\bf e.}~A deployable metamaterial undergoing dramatic, but controlled, shape changes using folding constraints and minimal actuation, adapted from Liu,~et~al.~Sci.~Robot~(2021).
    {\bf f.}~A colloidal diamond exemplifies self-assembly at the micron scale, adapted from He,~et~al.~Nature~(2020).
    }
\end{figure}

\noindent Soft metamaterials are composite structures whose collective mechanical properties go beyond those of their individual constituents~\cite{Bertoldi2017}.
For example, auxetic materials, which contract in all directions when squeezed, can be engineered via appropriate arrangements of flexible bonds.
\emph{Topological} metamaterials are those that use notions from topology (such as winding numbers or other topological invariants) to ensure the existence of a particular feature, typically a localized deformation mechanism or vibrational mode~\cite{Bertoldi2017}.  
For instance, the lattice in Fig.~\ref{Fig1}a is rigid everywhere, except near the colored defect where there is a soft mechanical mode protected by topology. Beyond materials with fixed properties, \emph{programmable} metamaterials use elastic multistability to encode specialized responses to external actuation.
In Fig.~\ref{Fig1}b, a pattern (e.g. a smiley face) is programmed to appear when the metamaterial is compressed.
\emph{Trainable} metamaterials turn programmability into a process that emerges from a sequence of experiences~\cite{Keim2019Memory}. 
For instance, the elasticity of the network in Fig.~\ref{Fig1}c is tuned through a process called directed aging, during which bonds are strengthened and weakened in response to external stress. The individual building blocks can also have an internal source of energy~\cite{Shankar2020}. This is the case for the \emph{active} and \emph{robotic} metamaterial shown in Fig.~\ref{Fig1}d, in which a wall of motorized hinges steers the outgoing direction of a projectile. 
In such active media, the familiar symmetries and conservation laws of passive matter need to be revisited to write down effective elastic models~\cite{Scheibner2019,Fruchart2020}. The concepts behind soft metamaterials, such as topology or continuum mechanics, often apply across length scales. For example, the world's smallest origami bird in Fig.~\ref{Fig1}e is a \emph{deployable} metamaterial that uses minimal actuation along with geometric constraints to guide its global shape change. One can even envision metamaterials that build themselves - a possibility particularly interesting at small scales - like the \emph{self-assembled} colloidal structure in Fig.~\ref{Fig1}f that mimics the atomic arrangement of diamond.

\endcountem 
\medskip\medskip\medskip

\medskip\medskip\medskip
\medskip\medskip\medskip

\clearpage 

\noindent \textbf{Current and Future Challenges.}
\countem 

\noindent 
The design of trainable, deployable, or self-assembled metamaterials exploits dynamic pathways.
Yet, the functionalities targeted by such dynamic pathways are often encoded in static structures (e.g. buckled or self-assembled states) with desired mechanical properties (like rigidity or a certain vibrational spectrum).
Programmable active or robotic elements embedded in the metamaterial allow us to envision inherently dynamical functionalities such as self-sustained motion and learning.

\begin{figure}
    \centering
    \includegraphics[width=0.7 \textwidth ]{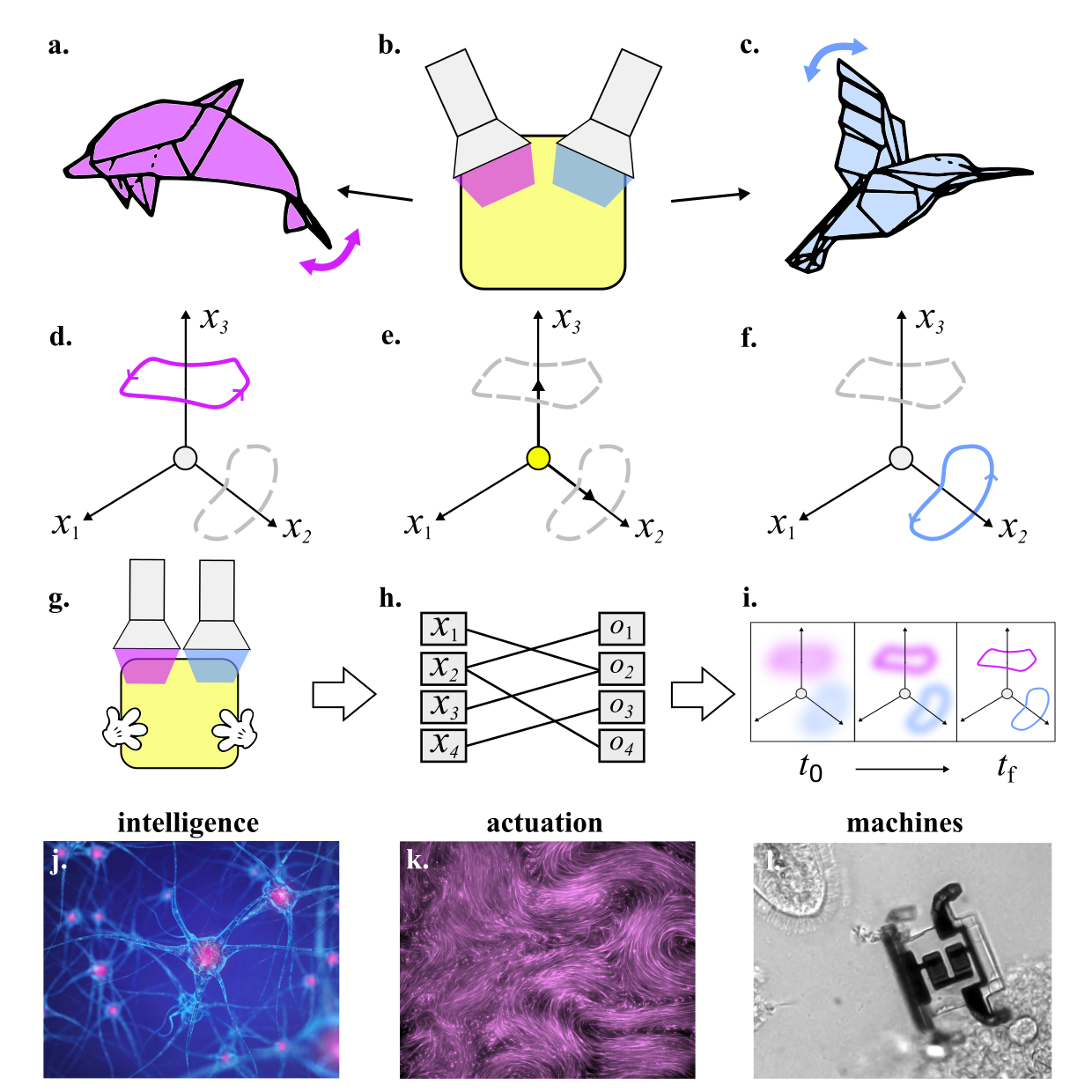}
    \caption{
    \label{Fig2}
    \textbf{Adaptive and intelligent soft metamaterials.}~{\bf a-c.} Schematic picture of a dynamic and intelligent soft metamaterial.  When exposed to a stimulus (represented by light), the metamaterial performs a computation, actuates, and reaches a dynamical steady-state that serves a function such as swimming or flying. {\bf d-f.} From the point of view of dynamical systems, these dynamical steady-states are attractors (such as limit cycles). Functionalities emerge by engineering suitable attractors and bifurcations.   
    {\bf g-i.} An adaptive metamaterial performs computation and learns. For example, physical stimulation (panel g) is processed by the material, which acts as a distributed machine-learning algorithm (panel h). The outputs of this process are, for instance, adapted dynamical states (panel i). 
    {\bf j.} Neurons are biology's solution for intelligent distributed materials, courtesy~\emph{nobeastsofierce}.
    {\bf k.} Physicochemical processes such as ATP consumption by kinesin motors power large-scale motion like active nematic flows, adapted from Duclos,~et~al.~Science~(2020).   
    {\bf l.} Micromachines the size of a single-cell organism provide a platform for intelligent and adaptive soft metamaterials, adapted from Miskin~et~al.~Nature~(2020).   
    }
\end{figure}

\emph{Metamaterials as dynamical systems.} 
A first challenge consists in engineering functionalities rooted in the nonlinear dynamics of the metamaterial. Figure~\ref{Fig2}a-c illustrates this objective: when stimulated, an initially homogeneous and undifferentiated piece of metamaterial transitions to a deployed state (dolfin- or bird-like states in panels a and c) that depends on the stimulus (purple and blue light in panel b). Crucially, this deployed state is not a static fixed point but a dynamic state that allows the metamaterial to act like a robot: it repeatedly performs actions that do work on their surroundings (swimming and walking).
Formally, these dynamic states are limit cycles or more complex attractors in configuration space.
This is shown in Fig.~\ref{Fig2}d-f, 
in which each axis represents a dynamical degree of freedom. 
Reaching this goal requires designing dynamical states that perform the required tasks, ensuring that they can be dynamically reached, and implementing them at the appropriate scales.

\emph{Metamaterials as computers.} 
A second challenge consists in conceptualizing and implementing some amount of intelligence in metamaterials: performing simple calculations in situ, learning and performing a variety of tasks from the same undifferentiated form, and adapting to external fluctuations.
Materials that change over time naturally store representations of their previous experiences~\cite{Keim2019Memory}. A key objective consists in creating materials that can manipulate and act upon this information in desired ways. Figure~\ref{Fig2}g-i illustrates the end goal: as the metamaterial is externally deformed under a colored light, it learns to perform a desired motion.
This level of complexity could be achieved by metamaterials that effectively perform simple machine learning algorithms using only distributed physical processes.  Such a form of distributed intelligence would lead to materials that are truly adaptive, like army ants forming and maintaining a bridge with their bodies.

\endcountem

\clearpage 

\noindent \textbf{Advances in Science and Technology to Meet Challenges.}
\countem 

\noindent 
\emph{Fabrication and programming.} Three main goals include miniaturizing and mass producing small elements that can be combined into a material-like structures; {decentralizing control} to remove the need of an external computer; and {integrating computation and actuation} so that part of the computation is delegated to the underlying physics.
To achieve these goals, advances in manufacturing will need to be combined with the design of decentralized algorithms tailored to the limited abilities of the building blocks.
Ideally, these building blocks should be able to repair and replicate themselves.
In that respect, biology is a powerful source of inspiration: neural tissues (Fig.~\ref{Fig2}j) are able to form and process complex representations,  
while muscle tissues
produce large-scale deformations (Fig.~\ref{Fig2}k). 
Integrating both the computation and actuation into a single metamaterial could be done using bio-inspired systems, such as engineered DNA molecules capable of implementing algorithms~\cite{Woods2019Diverse} and acting as robots that perform complex tasks~\cite{Anupama2017cargo}; or using micro-robots: small, electronically integrated machines, which can now be manufactured by the millions at the micron scale~\cite{Miskin2020electronically}.

\emph{Effective theories of adaptive dynamics.} 
Adaptive metamaterials are complex: they can adapt over long time scales while sustaining continued microscopic motions at very short time scales. 
This presents a challenge in describing the intermediate time scales relevant to material functionalities.
Hence, describing adaptive metamaterials requires going beyond approaches relying on the simple separation of time scales characteristic of hydrodynamic theories. 
Model reduction techniques developed in the context of statistics and dynamical systems, such as dimensionality reduction and invariant manifold reduction, can serve as a basis for developing such effective theories of adaptive dynamics.
In particular, these effective theories must rise to the challenge of describing phenomena such as physical learning~\cite{Stern2020Continual} and back propagation~\cite{Wright2021Deep}, which occur in non-linear glassy model systems. 
When these learning processes occur, the material effectively produces a representation of its past experiences which 
can be leveraged to perform some of the computations necessary to achieve desired functionalities.

\emph{Machine learning for material intelligence and design.} 
Finally,  machine learning can be used to accelerate and automate the inverse problem of design: how to go from target functionalities to the basic building blocks. For instance, machine learning can help establish relationships between tunable physical parameters and the dynamical attractors that define the dynamic functionalities of adaptive soft materials. Moreover, machine learning can facilitate dimensionality reduction techniques for forming effective theories, thereby offering the possibility of starting directly from experimental data.

\endcountem

\clearpage 

\textbf{Concluding Remarks.}
\countem 
Adaptive and intelligent soft metamaterials raise the prospect of merging matter with computers, and hold promise for applications ranging from medical science to space exploration. Reaching these promises will require tackling conceptual and practical challenges in fabrication as well as answering fundamental questions about the nature of representation and learning.

\endcountem

%\bibliographystyle{unsrt}
%\bibliography{biblio}

\end{document}